# Origami Cubes with One-DOF Rigid and Flat Foldability

Yuanqing Gu[1,2], Yan Chen[1,2*]

1. School of Mechanical Engineering, Tianjin University, Tianjin 300072, China

2. Key Laboratory of Mechanism Theory and Equipment Design of Ministry of Education, Tianjin University, Tianjin 300072, China

* Corresponding Author.

   Email: yan_chen@tju.edu.cn, ORCID ID: 0000-0002-4742-6944

**Abstract**

Rigid origami is a branch of origami with great potential in engineering applications to deal with rigid-panel folding. One of the challenges is to compactly fold the polyhedra made from rigid facets with a single degree of freedom. In this paper, we present a new method to design origami cubes with three fundamental characteristics, rigid foldability, flat foldability and one degree of freedom (DOF). A total of four cases of crease patterns that enable origami cubes with distinct folding performances have been proposed with all possible layouts of the diagonal creases on the square facets of origami cubes. Moreover, based on the kinematic equivalence between the rigid origami and the spherical linkages, the corresponding spherical linkage loops are introduced and analysed to reveal the motion properties of the origami cubes. The newly found method can be readily utilized to design deployable structures for various engineering applications including cube-shaped cartons, small satellites, containers, etc.

**Keywords:**

Rigid origami; Flat foldability; Origami cube; Spherical linkage loop.

# 1. Introduction

Origami, the oriental art of paper folding, has attracted more and more attention from mathematicians, scientists and engineers (Peraza-Hernandez et al., 2014) as it can create three-dimensional (3D) structures from two-dimensional (2D) flat sheets of paper through a process of folding along predetermined creases (Lv et al., 2014). Many foldable structures inspired by origami have been widely adopted in aerospace, civil engineering and transportation industries due to its extraordinary ability of folding a large structure into a compact size, such as solar arrays (Miura, 1994 and 2009), satellite antenna reflectors (Zirbel et al., 2013), deployable origami tents (Liu et al., 2016), and so on.

Rigid origami is a subset of origami that remains foldable even when its facets and creases are replaced by rigid panels and hinges (Wang and Chen, 2011), i.e., its facets revolve around designed creases without any deformation during the continuous folding process. In kinematics, the facets and creases of rigid origami can also be regarded as rigid links and revolute joints (Dai and Jones, 1999), so, the creases that intersect at one vertex can be modeled as a spherical linkage and the crease pattern as a network of spherical linkages (Demaine and O'Rourke, 2007). Furthermore, several approaches have been proposed to analyse the rigid foldability of origami patterns including numerical algorithms (Nojima, 2002), quaternions and dual quaternions (Wu and You, 2010), and matrix methods (Tachi, 2009).

Meanwhile, flat foldability is another important property for an origami pattern in both theory investigation and practical application, which refers to the capability to fold an origami pattern compactly into a flat overlapped sheet (Tachi, 2010). Origami structures with flat foldability provide a great convenience in practice in terms of portability and storage space. Theoretically, Hull (2013) studied the flat foldability of origami patterns through the assignment of mountain and valley creases with mathematical theory. Schneider (2004) provided the conditions for flat foldability of an arbitrary unsigned origami pattern. (Tachi, 2010) developed a technique to generalize the geometry of cylindrical surface using bidirectionally flat-foldable planar quadrilateral mesh.

Apart from flat foldability, engineers also work on how to fold a patterned sheet into a box-shaped carton (Dai and Cannella, 2008). Yet, there is little work on how to fold cartons compactly into a flat configuration. Mathematicians concur that it is impossible to fold a sealed box rigidly (Connelly et al., 1997). Hence, two rigid folding schemes for cubical bags without top cover have been proposed by Balkcom et al. (2014) and the solution for rigid and flat folding of tall bags has been found by Wu and You (2010). There has been no approach to make origami cubes with all six facets that satisfy rigid foldability, flat foldability and one degree of freedom (DOF) at the same time. Therefore, in this paper, we explore the possibility to solve this challenge by introducing one diagonal cut on the top facet of the cube.

The layout of the paper is as follows. Section 2 proposes a basic crease pattern of the rigid and flat foldable origami cube and analyses kinematic behavior of the corresponding $4R$-$5R$-$4R$-$5R$ spherical linkage loop. By exploring the layouts of diagonal creases on the rigid facets of origami cube, Section 3 presents three more effective crease patterns. Finally, the conclusion and further discussion are given in Section 4.

## 2. The $4R$-$5R$-$4R$-$5R$ origami cube

The well-known non-rigid but flat foldable crease pattern of the origami cube is shown in Fig. 1(a) (Guest and Pellegrino, 1994). In this pattern, in addition to the inherent edge creases of the cube, there are four diagonal creases ($AB'$, $BC'$, $CD'$ and $DA'$) on the four vertical facets of the cube, respectively. Although this origami cube is flat foldable in the vertical direction, the deformation on facets exists during the folding process. Here, we propose a modified crease pattern, as shown in Fig. 1(b), by adding two extra diagonal creases ($BD$, $A'C'$) and one incision ($B'D'$) to the crease pattern in Fig. 1(a).

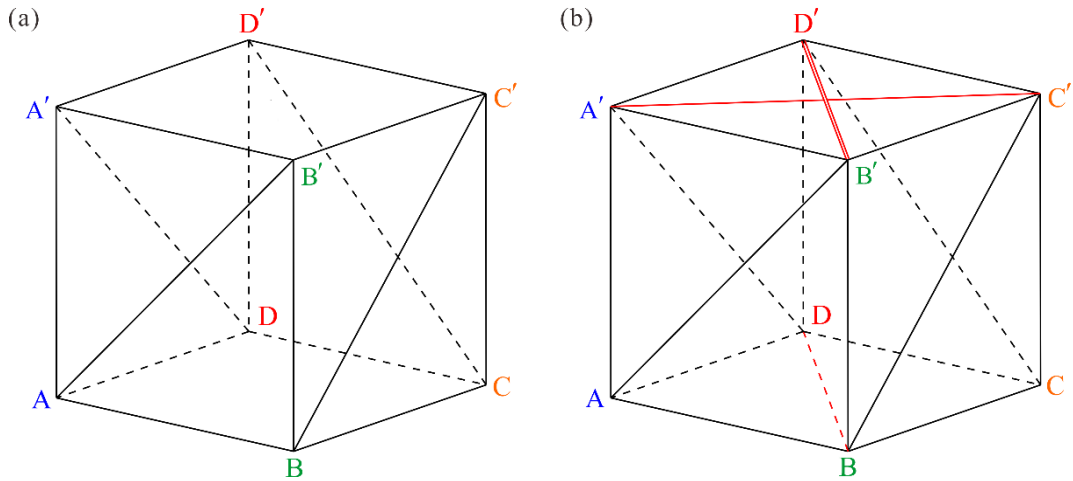

Fig.1. (a) The non-rigid but flat foldable crease pattern. (b) The rigid and flat foldable crease pattern.

On the bottom of the origami cube, there are four creases at vertices A and C, and five creases at vertices B and D, which can be modelled as spherical 4$R$ (S4$R$) and 5$R$ (S5$R$) linkages, respectively. Meanwhile, each pair of two adjacent spherical linkages shares one revolute joint and the two S5$R$ linkages share a common joint on crease BD. Thus, they form the 4$R$-5$R$-4$R$-5$R$ (from vertex A to vertex D) spherical linkage loop as shown in Fig.2. In this loop, two S4$R$ linkages are rotationally symmetric with respect to the centerline of the cube, and so do two S5$R$ linkages.

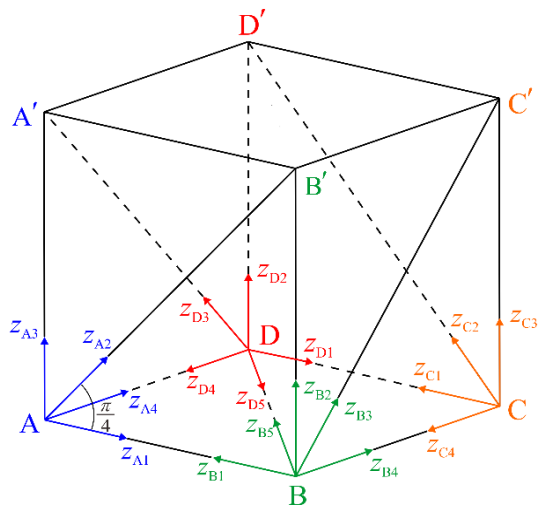

Fig.2. The 4$R$-5$R$-4$R$-5$R$ spherical linkage loop.

Kinematic analysis of this spherical linkage loop can be carried out based on the matrix method with the Denavit and Hartenberg notations (Denavit, 1955). Fig.3a shows a portion of a spherical linkage, whose all rotating axes intersect at one point in space. Here, $z_i$ is along the axis of revolute joint $i$, and $x_i$ is along the direction of common normal from $z_{i-1}$ to $z_i$ by the right-hand rule. The geometrical parameter $\alpha_{i(i+1)}$ represents the angle between $z_i$ and $z_{i+1}$ about the axis $x_{i+1}$. The kinematic variable $\theta_i$ is defined as the rotation angle from $x_i$ to $x_{i+1}$, positive along the axis $z_i$. In general, for a single closed loop linkage consisting of $i$ links, the closure equation is

$$Q_{21}Q_{32}Q_{43}\cdots Q_{i(i-1)}Q_{1i} = I, \tag{1}$$

where the transformation matrix $Q_{(i+1)i}$ is

$$Q_{(i+1)i} = \begin{bmatrix} \cos\theta_i & -\cos\alpha_{i(i+1)}\sin\theta_i & \sin\alpha_{i(i+1)}\sin\theta_i \\ \sin\theta_i & \cos\alpha_{i(i+1)}\cos\theta_i & -\sin\alpha_{i(i+1)}\cos\theta_i \\ 0 & \sin\alpha_{i(i+1)} & \cos\alpha_{i(i+1)} \end{bmatrix}, \tag{2}$$

and the inverse transformation $Q_{i(i+1)}$ can be expressed as

$$Q_{i(i+1)} = \begin{bmatrix} \cos\theta_i & \sin\theta_i & 0 \\ -\cos\alpha_{i(i+1)}\sin\theta_i & \cos\alpha_{i(i+1)}\cos\theta_i & \sin\alpha_{i(i+1)} \\ \sin\alpha_{i(i+1)}\sin\theta_i & -\sin\alpha_{i(i+1)}\cos\theta_i & \cos\alpha_{i(i+1)} \end{bmatrix}. \tag{3}$$

According to the D-H matrix method, let us present the kinematics of four spherical linkages at vertices A, B, C and D, respectively. At vertex A, see Fig.3(b), the geometrical parameters $\alpha_{Ai(i+1)}$ ($i$ =1, 2, 3, 4, when $i+1>4$, it is replaced by 1) of the S4R linkage are

$$\alpha_{A12}=\alpha_{A23}=\frac{\pi}{4}, \quad \alpha_{A34}=\alpha_{A41}=\frac{\pi}{2}, \tag{4}$$

and the closure equation in Eq. (1) can be rewritten as

$$Q_{23}Q_{12} = Q_{43}Q_{14}. \tag{5}$$

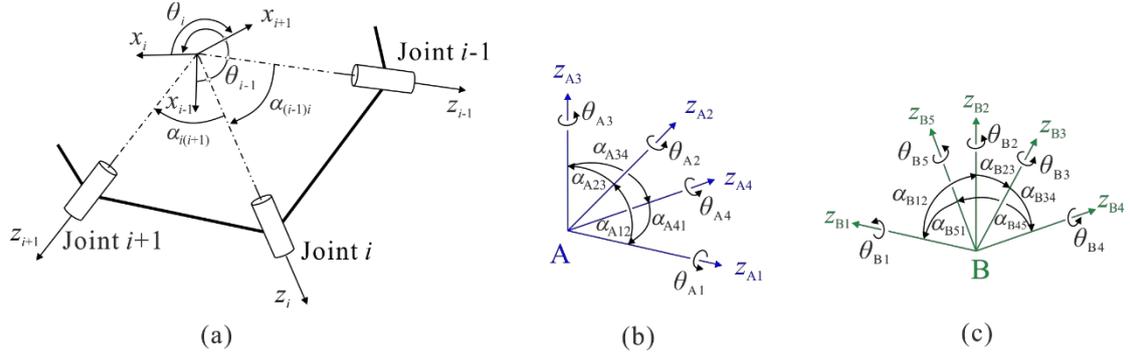

Fig.3. The D-H notations of (a) a portion of a spherical linkage, (b) S4R linkage A and (c) S5R linkage B.

Substituting the geometrical parameters into Eq. (5), the kinematic relationships can be derived as

$$\tan\frac{\theta_{A2}}{2}=\frac{1}{\sqrt{2}\tan\theta_{A1}},\quad \theta_{A3}=\theta_{A1},\quad \tan\frac{\theta_{A4}}{2}=\frac{1}{\sin\theta_{A1}}, \qquad (6)$$

where $\theta_{A1}$ is taken as the input kinematic variable for one-DOF S4R linkage.

The geometrical parameters of the two S4R linkages at vertices A and C are identical. Hence when $\theta_{C1}$ is the input kinematic variable, the kinematics of the S4R linkage at vertex C is

$$\tan\frac{\theta_{C2}}{2}=\frac{1}{\sqrt{2}\tan\theta_{C1}},\quad \theta_{C3}=\theta_{C1},\quad \tan\frac{\theta_{C4}}{2}=\frac{1}{\sin\theta_{C1}}. \qquad (7)$$

For the two-DOF S5R linkage at vertex B, see Fig.3(c), the geometrical parameters $\alpha_{Bi(i+1)}$ ($i$=1, 2, 3, 4, 5, when $i$+1>5, it is replaced by 1) are

$$\alpha_{B12}=\frac{\pi}{2},\quad \alpha_{B23}=\alpha_{B34}=\alpha_{B45}=\alpha_{B51}=\frac{\pi}{4}. \qquad (8)$$

The closure equation (1) can be rewritten as

$$\boldsymbol{Q}_{34}\boldsymbol{Q}_{23}=\boldsymbol{Q}_{54}\boldsymbol{Q}_{15}\boldsymbol{Q}_{21}. \qquad (9)$$

Substituting Eq. (8) into Eq. (9), the following kinematic relationships can be obtained

$$\tan\frac{\theta_{B5}}{2} = \frac{\left(\begin{array}{c}-\frac{\sqrt{2}}{2}\sin(\theta_{B1}+\theta_{B4})-\frac{\sqrt{2}}{2}\sin\theta_{B1} \\ +\sqrt{\frac{1}{2}\sin^2\theta_{B4}-\cos^2\theta_{B1}+\cos\theta_{B1}\cos\theta_{B4}-\cos\theta_{B1}+\cos\theta_{B4}}\end{array}\right)}{\cos(\theta_{B1}+\theta_{B4})-1}, \quad (10a)$$

$$\cos\theta_{B2} = \frac{1}{2}\cos\theta_{B4}\cos\theta_{B5} - \frac{\sqrt{2}}{2}\sin\theta_{B4}\sin\theta_{B5} + \frac{1}{2}\cos\theta_{B4} + \frac{1}{2}\cos\theta_{B5} - \frac{1}{2}, \quad (10b)$$

$$\cos\theta_{B3} = -\sqrt{2}\sin\theta_{B1}\sin\theta_{B5} + \cos\theta_{B1}\cos\theta_{B5} + \cos\theta_{B1} + 1. \quad (10c)$$

where $\theta_{B1}$ and $\theta_{B4}$ are taken as two needed inputs as S5R linkage is of two-DOF.

Since the two S5R linkages at vertices B and D have identical geometrical parameters, we can have the following kinematic relationships for linkage D by taking $\theta_{D1}$ and $\theta_{D4}$ as inputs

$$\tan\frac{\theta_{D5}}{2} = \frac{\left(\begin{array}{c}-\frac{\sqrt{2}}{2}\sin(\theta_{D1}+\theta_{D4})-\frac{\sqrt{2}}{2}\sin\theta_{D1} \\ +\sqrt{\frac{1}{2}\sin^2\theta_{D4}-\cos^2\theta_{D1}+\cos\theta_{D1}\cos\theta_{D4}-\cos\theta_{D1}+\cos\theta_{D4}}\end{array}\right)}{\cos(\theta_{D1}+\theta_{D4})-1}, \quad (11a)$$

$$\cos\theta_{D2} = \frac{1}{2}\cos\theta_{D4}\cos\theta_{D5} - \frac{\sqrt{2}}{2}\sin\theta_{D4}\sin\theta_{D5} + \frac{1}{2}\cos\theta_{D4} + \frac{1}{2}\cos\theta_{D5} - \frac{1}{2}, \quad (11b)$$

$$\cos\theta_{D3} = -\sqrt{2}\sin\theta_{D1}\sin\theta_{D5} + \cos\theta_{D1}\cos\theta_{D5} + \cos\theta_{D1} + 1. \quad (11c)$$

In the coordinate system shown in Fig.2, each pair of adjacent spherical linkages share a common revolute joint in the 4R-5R-4R-5R spherical linkage loop. Then, we can obtain the following relationships of kinematic variables

$$\theta_{A1} = -\theta_{B1}, \quad \theta_{B4} = -\theta_{C4}, \quad \theta_{C1} = -\theta_{D1}, \quad \theta_{D4} = -\theta_{A4}, \quad (12a)$$

as the two S5R linkages share a common joint BD, we have

$$\theta_{B5} = \theta_{D5}. \quad (12b)$$

By combining Eqs. (6-7) and (10-12), we can obtain

$$\theta_{A1}=\theta_{C1}. \tag{13}$$

For the 4R-5R-4R-5R spherical linkage loop,

$$\tan\frac{\theta_{A2}}{2}=\frac{1}{\sqrt{2}\tan\theta_{A1}}, \quad \theta_{A3}=\theta_{A1}, \quad \tan\frac{\theta_{A4}}{2}=\frac{1}{\sin\theta_{A1}},$$

$$\theta_{B1}=-\theta_{A1}, \quad \theta_{B4}=-\theta_{A4},$$

$$\tan\frac{\theta_{B5}}{2}=\frac{\dfrac{\sqrt{2}}{2}\sin(\theta_{A1}+\theta_{A4})+\dfrac{\sqrt{2}}{2}\sin\theta_{A1}+\sqrt{\dfrac{1}{2}\sin^2\theta_{A4}-\cos^2\theta_{A1}+\cos\theta_{A1}\cos\theta_{A4}-\cos\theta_{A1}+\cos\theta_{A4}}}{\cos(\theta_{A1}+\theta_{A4})-1},$$

$$\cos\theta_{B2}=\frac{1}{2}\cos\theta_{A4}\cos\theta_{B5}+\frac{\sqrt{2}}{2}\sin\theta_{A4}\sin\theta_{B5}+\frac{1}{2}\cos\theta_{A4}+\frac{1}{2}\cos\theta_{B5}-\frac{1}{2},$$

$$\cos\theta_{B3}=\sqrt{2}\sin\theta_{A1}\sin\theta_{B5}+\cos\theta_{A1}\cos\theta_{B5}+\cos\theta_{A1}+1.$$

$$\theta_{Ci}=\theta_{Ai} \quad (i=1, 2, 3, 4),$$

$$\theta_{Di}=\theta_{Bi} \quad (i=1, 2, 3, 4, 5). \tag{14}$$

Hence, if $\theta_{A1}$ is given as only one input kinematic variable of the 4R-5R-4R-5R spherical linkage loop, the other seventeen outputs can be obtained and the motion of the origami cube can be determined, i.e., the origami cube has one DOF and rigid foldability.

Generally, in origami study, dihedral angles are preferred to directly represent the folding process. In this origami cube, the relationships between the dihedral angles and kinematic variables are

$$\varphi_{A1}=\pi+\theta_{A1}, \quad \varphi_{A2}=\pi-\theta_{A2}, \quad \varphi_{A3}=\pi+\theta_{A3}, \quad \varphi_{A4}=\pi+\theta_{A4}, \tag{15a}$$

$$\varphi_{B1}=\pi-\theta_{B1}, \quad \varphi_{B2}=\pi-\theta_{B2}, \quad \varphi_{B3}=\pi-\theta_{B3}, \quad \varphi_{B4}=\pi-\theta_{B4}, \quad \varphi_{B5}=\pi+\theta_{B5}, \tag{15b}$$

$$\varphi_{C1}=\pi+\theta_{C1}, \quad \varphi_{C2}=\pi-\theta_{C2}, \quad \varphi_{C3}=\pi+\theta_{C3}, \quad \varphi_{C4}=\pi+\theta_{C4}, \tag{15c}$$

$$\varphi_{D1}=\pi-\theta_{D1}, \quad \varphi_{D2}=\pi-\theta_{D2}, \quad \varphi_{D3}=\pi-\theta_{D3}, \quad \varphi_{D4}=\pi-\theta_{D4}, \quad \varphi_{D5}=\pi+\theta_{D5}. \tag{15d}$$

Substituting Eq. (15) to Eq. (14), the following equations can be obtained

$$\tan\frac{\varphi_{A2}}{2}=\sqrt{2}\tan\varphi_{A1},\quad \varphi_{A3}=\varphi_{A1},\quad \tan\frac{\varphi_{A4}}{2}=\sin\varphi_{A1},$$

$$\varphi_{B1}=\varphi_{A1},\quad \varphi_{B4}=\varphi_{A4},$$

$$\tan\frac{\varphi_{B5}}{2}=\frac{-\cos(\varphi_{A1}+\varphi_{A4})+1}{\dfrac{\sqrt{2}}{2}\sin(\varphi_{A1}+\varphi_{A4})-\dfrac{\sqrt{2}}{2}\sin\varphi_{A1}}{+\sqrt{\dfrac{1}{2}\sin^{2}\varphi_{A4}-\cos^{2}\varphi_{A1}+\cos\varphi_{A1}\cos\varphi_{A4}+\cos\varphi_{A1}-\cos\varphi_{A4}}},$$

$$\cos\varphi_{B2}=-\frac{1}{2}\cos\varphi_{A4}\cos\varphi_{B5}-\frac{\sqrt{2}}{2}\sin\varphi_{A4}\sin\varphi_{B5}+\frac{1}{2}\cos\varphi_{A4}+\frac{1}{2}\cos\varphi_{B5}+\frac{1}{2},$$

$$\cos\varphi_{B3}=-\sqrt{2}\sin\varphi_{A1}\sin\varphi_{B5}-\cos\varphi_{A1}\cos\varphi_{B5}+\cos\varphi_{A1}-1,$$

$$\varphi_{Ci}=\varphi_{Ai}\quad (i=1,2,3,4),$$

$$\varphi_{Di}=\varphi_{Bi}\quad (i=1,2,3,4,5), \tag{16}$$

where $\varphi_{A1}$ is the only input dihedral angle. Fig.4 shows the input-output curves of dihedral angles and motion sequences of the origami cube, which is folded into a planar square. A physical model was also fabricated to show the folding process as shown in Fig. 5, and its deployed and folded configurations are a cube and a planar square, respectively. During the folding process, it presents the rotational symmetry. Actually, the 4R-5R-4R-5R spherical linkage loop at the bottom of the origami cube drives the structure to realise the folding performance with one DOF.

As for the crease pattern on the top facet of the origami cube, the cut is made along diagonal $B'D'$, the crease is along $A'C'$, so vertices $A'$ and $C'$ form two S5R linkages. Take vertex C as an example, five creases are $C'C$, $C'B$, $C'B'$, $C'E$ and $C'D'$, in which the rotations on creases $C'C$ and $C'B$ are determined by the motion at vertices C and B respectively. Thus, the two creases are the inputs of the S5R linkage at vertex $C'$. Similarly, the motion at vertex $A'$ is also determined by the inputs from vertices A and D. Hence the origami cube with top facets is still of one DOF.

However, during the folding, creases $A'E$ and $C'E$ interfere with creases $AB'$ and $CD'$ respectively. We have added two extra creases, $A'F$ and $C'G$, to provide flexibility to the top facets to avoid the interference. In this case, the vertices

A′ and C′ are under-actuated. Yet, the whole cube is still exhibiting the one-DOF rigid and flat foldability as shown in Fig.5.

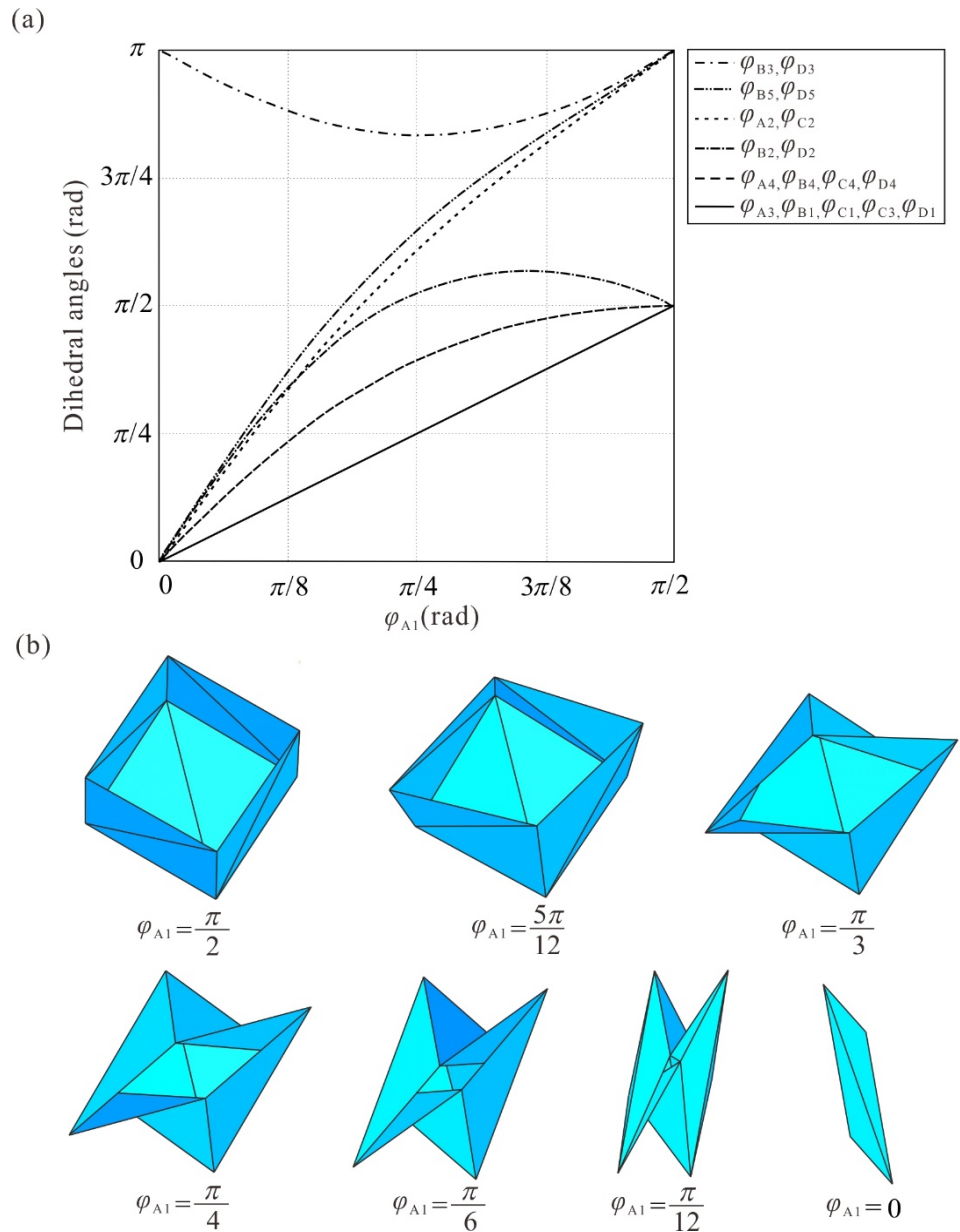

Fig.4. (a) The input-output curves of dihedral angles and (b) motion sequences of the 4R-5R-4R-5R origami cube.

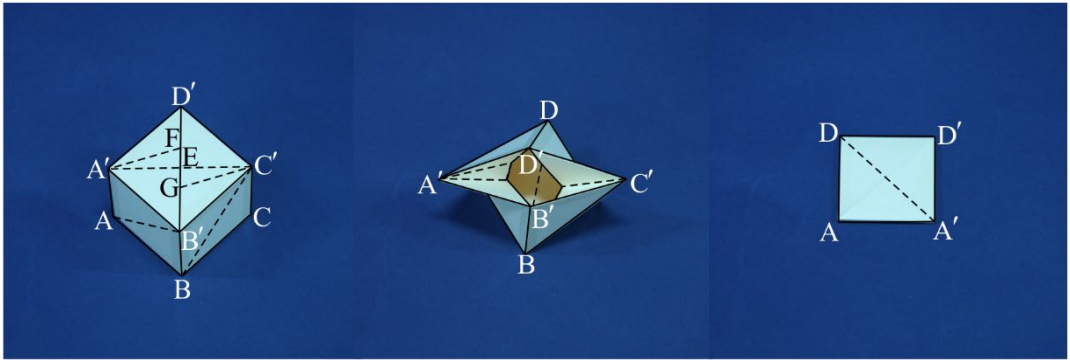

Fig. 5. The folding process of the 4*R*-5*R*-4*R*-5*R* origami cube.

## 3. The effective crease patterns of one-DOF rigid and flat foldable origami cubes

For the cubical structure in Fig 2, there are five diagonal creases. Next, we are going to explore all the possible schemes to realise the rigid and flat folding performance with one DOF by studying the distribution of these five diagonal creases. For convenience, planar nets of the cube without the top surface are shown in Fig.6, in which the central square is the bottom facet of the cube, and diagonals are the corresponding creases, to avoid duplication, it is defined by the red line on the central square. The remaining four diagonal creases on each planar net are shown by the blue line. Totally, there are sixteen (=$2^4$) possible distributions of the diagonal creases, each of which has two possible directions. Considering rotational and flipping symmetry, patterns ① and ② are identical, named as case 1, and patterns ③-⑥ as case 2, patterns ⑦ and ⑧ as case 3, pattern ⑨ as case 4, while patterns ⑩-⑯ form a tetrahedron with three adjacent grey triangles when folding into a cube, which make the cube lose the foldability.

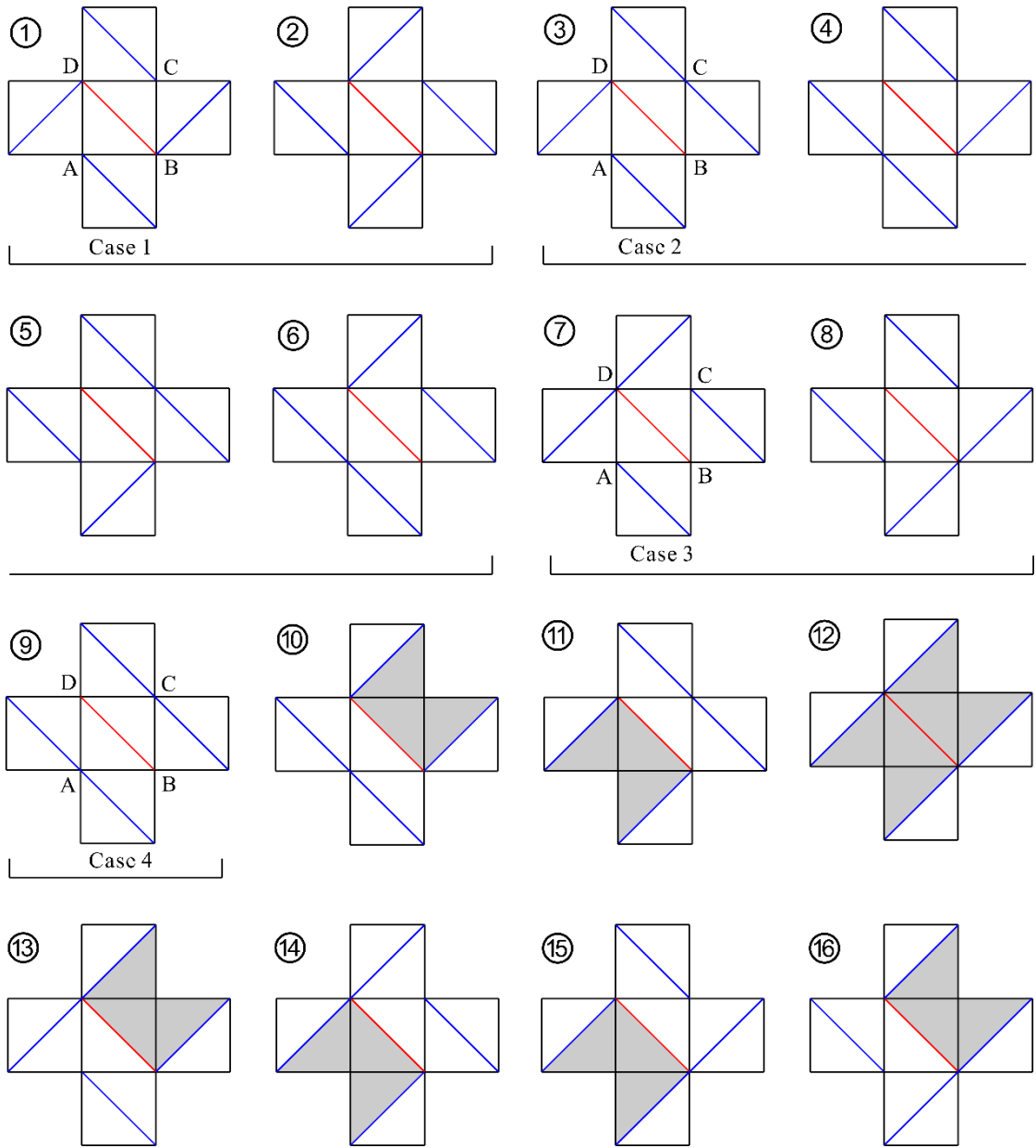

Fig.6. Planar nets of the cube without the top surface.

Case 1 is, in fact, the $4R$-$5R$-$4R$-$5R$ spherical linkage loop, which has been analysed in detail in section 2. Although patterns ① and ② can both be regarded as the $4R$-$5R$-$4R$-$5R$ spherical linkage loop, one difference is that the origami cube with pattern ① has an anti-clockwise folding performance (shown in Section 2, Fig.5) about the cube's vertical central line, while the origami cube with pattern ② is folded clockwise.

In case 2, patterns ③ and ④ can be regarded as the same pattern after 180° rotation, so do patterns ⑤ and ⑥. The crease pattern ③ is a 4R-4R-5R-5R (from vertex A to vertex D) spherical linkage loop as shown in Fig. 7. The geometric parameters of each spherical linkage are

$$\alpha_{A12}=\alpha_{A23}=\frac{\pi}{4}, \quad \alpha_{A34}=\alpha_{A41}=\frac{\pi}{2}, \tag{17a}$$

$$\alpha_{B12}=\alpha_{B23}=\frac{\pi}{4}, \quad \alpha_{B34}=\alpha_{B41}=\frac{\pi}{2}, \tag{17b}$$

$$\alpha_{C12}=\frac{\pi}{2}, \quad \alpha_{C23}=\alpha_{C34}=\alpha_{C45}=\alpha_{C51}=\frac{\pi}{4}, \tag{17c}$$

$$\alpha_{D12}=\frac{\pi}{2}, \quad \alpha_{D23}=\alpha_{D34}=\alpha_{D45}=\alpha_{D51}=\frac{\pi}{4}. \tag{17d}$$

As each pair of adjacent spherical linkages share a common joint, we can obtain the following relationships among dihedral angles,

$$\varphi_{A1}=\varphi_{B1}, \quad \varphi_{B3}=\varphi_{C2}, \quad \varphi_{C1}=\varphi_{D1}, \quad \varphi_{D4}=\varphi_{A4}, \quad \varphi_{B2}=\varphi_{D5}. \tag{18}$$

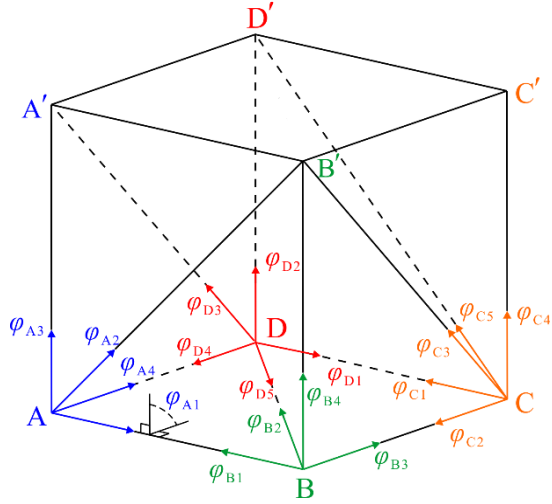

Fig.7. The representative crease pattern of case 2 and the 4R-4R-5R-5R spherical linkage loop.

The same approach described in Section 2 is utilized to analyse the kinematics in this loop. Taking $\varphi_{A1}$ as an input dihedral angle, the kinematics of two S4R linkages at vertices A and B can be determined. Subsequently, $\varphi_{D4}$ and $\varphi_{D5}$ can be obtained as inputs of the two-DOF S5R linkage at vertex D, while $\varphi_{D1}$ and $\varphi_{B3}$ are the inputs of S5R linkage at vertex C. Therefore, the 4R-4R-5R-5R spherical linkage loop has one DOF and its kinematic equations are

$$\tan\frac{\varphi_{A2}}{2}=\sqrt{2}\tan\varphi_{A1},\quad \varphi_{A3}=\varphi_{A1},\quad \tan\frac{\varphi_{A4}}{2}=\sin\varphi_{A1},$$

$$\varphi_{B1}=\varphi_{B3}=\varphi_{A1},\quad \tan\frac{\varphi_{B2}}{2}=\sqrt{2}\tan\varphi_{B1},\quad \tan\frac{\varphi_{B4}}{2}=\sin\varphi_{B1},$$

$$\varphi_{D4}=\varphi_{A4},\quad \varphi_{D5}=\varphi_{B2},$$

$$\tan\frac{\varphi_{D1}}{2}=\frac{-\frac{\sqrt{2}}{2}\sin\varphi_{D4}\sin\varphi_{D5}-\frac{1}{2}\cos\varphi_{D4}\cos\varphi_{D5}-\frac{1}{2}\cos\varphi_{D4}+\frac{1}{2}\cos\varphi_{D5}+\frac{1}{2}}{\left[\begin{array}{l}-\sin\varphi_{D4}\cos\varphi_{D5}+\frac{\sqrt{2}}{2}\cos\varphi_{D4}\sin\varphi_{D5}-\frac{\sqrt{2}}{2}\sin\varphi_{D5}\\-\sqrt{\left(-\frac{\sqrt{2}}{2}\sin\varphi_{D4}\sin\varphi_{D5}-\frac{1}{2}\cos\varphi_{D4}\cos\varphi_{D5}-\frac{1}{2}\cos\varphi_{D4}+\frac{1}{2}\cos\varphi_{D5}-\frac{1}{2}\right)^2}\\+\left(-\sin\varphi_{D4}\cos\varphi_{D5}+\frac{\sqrt{2}}{2}\cos\varphi_{D4}\sin\varphi_{D5}-\frac{\sqrt{2}}{2}\sin\varphi_{D5}\right)^2-1\end{array}\right]},$$

$$\cos\varphi_{D3}=-\sqrt{2}\sin\varphi_{D1}\sin\varphi_{D5}-\cos\varphi_{D1}\cos\varphi_{D5}+\cos\varphi_{D1}-1,$$

$$\cos\varphi_{D2}=-\frac{1}{2}\cos\varphi_{D4}\cos\varphi_{D5}-\frac{\sqrt{2}}{2}\sin\varphi_{D5}\sin\varphi_{D4}+\frac{1}{2}\cos\varphi_{D4}+\frac{1}{2}\cos\varphi_{D5}+\frac{1}{2},$$

$$\varphi_{C1}=\varphi_{D1},\quad \varphi_{C2}=\varphi_{B3},$$

$$\tan\frac{\varphi_{C3}}{2}=\frac{\cos\varphi_{C1}-1}{\left(\begin{array}{l}\frac{\sqrt{2}}{2}\sin\varphi_{C2}-\frac{\sqrt{2}}{2}\sin\varphi_{C1}\cos\varphi_{C2}\\-\sqrt{\frac{1}{2}\sin^2\varphi_{C1}\cos^2\varphi_{C2}-\left(\sin\varphi_{C1}\sin\varphi_{C2}-1\right)\left(\cos\varphi_{C1}+\cos\varphi_{C2}\right)}\\+\frac{1}{2}\sin^2\varphi_{C2}-\cos\left(\varphi_{C1}+\varphi_{C2}\right)-1\end{array}\right)},$$

$$\cos\varphi_{C5}=-\sqrt{2}\sin\varphi_{C2}\sin\varphi_{C3}-\cos\varphi_{C2}\cos\varphi_{C3}+\cos\varphi_{C2}-1,$$

$$\sin\varphi_{C4}=\sin\varphi_{C1}\cos\varphi_{C2}\cos\varphi_{C3}+\frac{\sqrt{2}}{2}\sin\varphi_{C1}\sin\varphi_{C2}\sin\varphi_{C3}-\frac{\sqrt{2}}{2}\cos\varphi_{C1}\sin\varphi_{C3}$$
$$-\sin\varphi_{C2}\cos\varphi_{C3}+\frac{\sqrt{2}}{2}\cos\varphi_{C2}\sin\varphi_{C3}. \quad (19)$$

The input-output curves of dihedral angles and motion sequences are shown in Fig.8. Although the geometric parameters of two S5R linkages are identical, their folding motions are distinct because of the different inputs. Hence, this origami cube has no special symmetric performance during folding but still can be rigid and flat foldable with one DOF, see Fig.9.

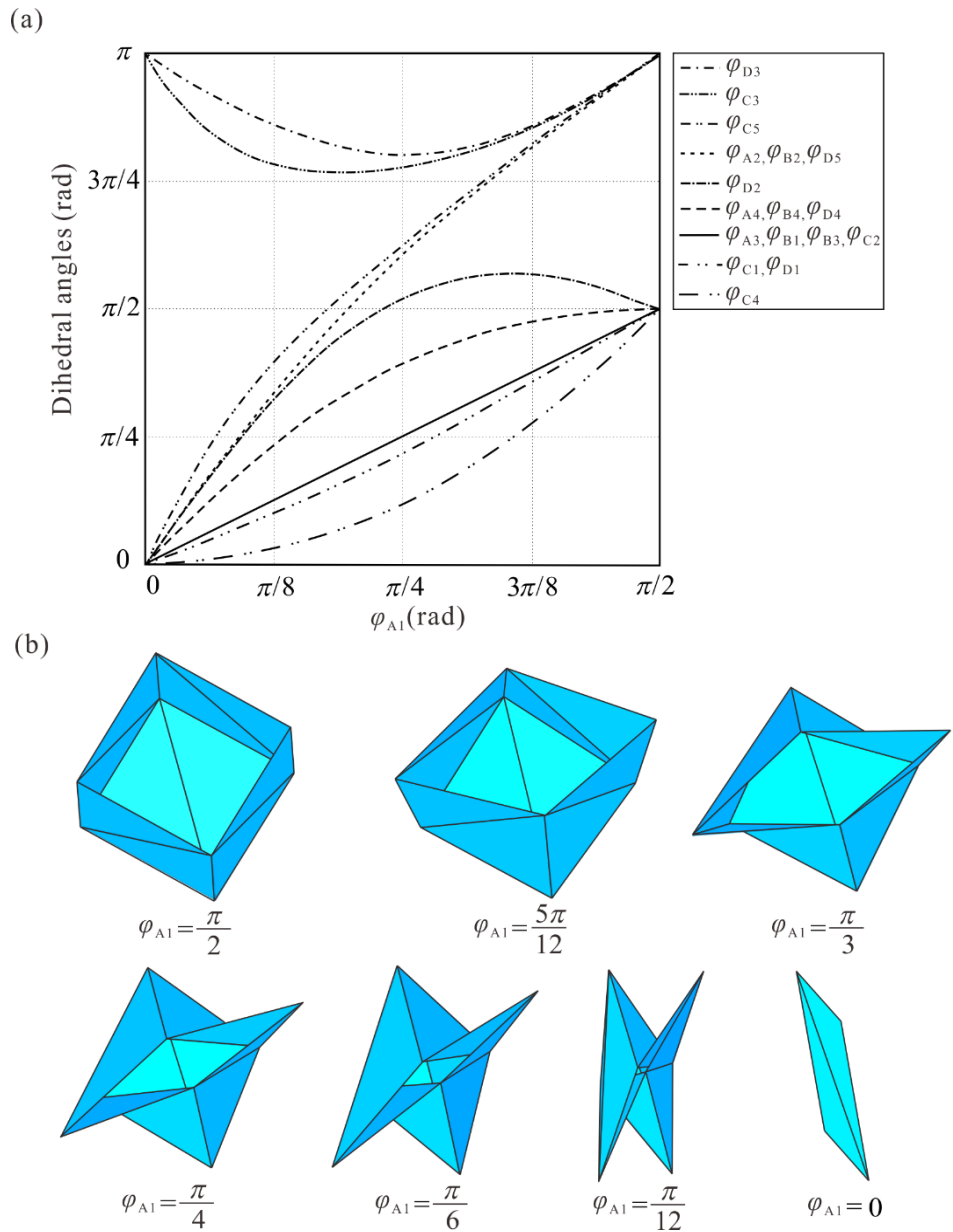

Fig.8. (a) The input-output curves of dihedral angles and (b) the motion sequences of the origami cube in case 2.

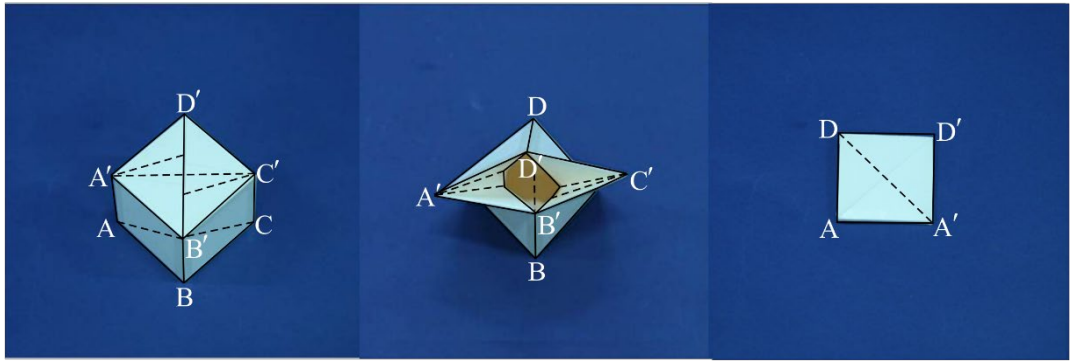

Fig. 9. The folding process of the 4R-4R-5R-5R origami cube.

In case 3, take pattern ⑦ as the example, which forms a 4R-4R-4R-6R (from vertex A to vertex D) spherical linkage loop (Fig.10). The geometric parameters of each spherical linkage are

$$\alpha_{A12}=\alpha_{A23}=\frac{\pi}{4},\ \alpha_{A34}=\alpha_{A41}=\frac{\pi}{2}, \tag{20a}$$

$$\alpha_{B12}=\alpha_{B23}=\frac{\pi}{4},\ \alpha_{B34}=\alpha_{B41}=\frac{\pi}{2}, \tag{20b}$$

$$\alpha_{C12}=\alpha_{C23}=\frac{\pi}{4},\ \alpha_{C34}=\alpha_{C41}=\frac{\pi}{2}, \tag{20c}$$

$$\alpha_{D12}=\alpha_{D23}=\alpha_{D34}=\alpha_{D45}=\alpha_{D56}=\alpha_{D61}=\frac{\pi}{4}. \tag{20d}$$

The kinematic relationships of each pair of adjacent spherical linkages can be expressed by dihedral angles as

$$\varphi_{A1}=\varphi_{B1},\ \varphi_{B3}=\varphi_{C3},\ \varphi_{C4}=\varphi_{D2},\ \varphi_{D6}=\varphi_{A4},\ \varphi_{B2}=\varphi_{D1}. \tag{21}$$

If $\varphi_{A1}$ is given as an input dihedral angle, the kinematic properties of three S4R linkages are determined. Meanwhile, three inputs ($\varphi_{D1}$, $\varphi_{D2}$ and $\varphi_{D6}$) required by three-DOF S6R linkage can also be obtained from $\varphi_{B2}$, $\varphi_{C4}$, $\varphi_{A4}$ respectively. Hence, as expressed in Eq. (22), the 4R-4R-4R-6R spherical linkage loop is one-DOF.

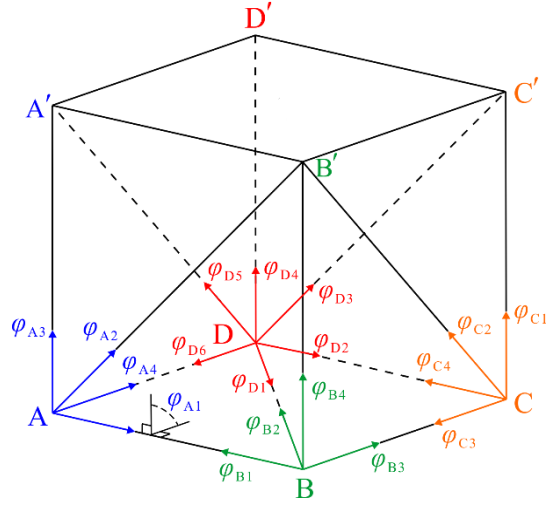

Fig.10. The representative crease pattern of case 3 and the 4R-4R-4R-6R spherical linkage loop.

$$\tan\frac{\varphi_{A2}}{2}=\sqrt{2}\tan\varphi_{A1},\ \varphi_{A3}=\varphi_{A1},\ \tan\frac{\varphi_{A4}}{2}=\sin\varphi_{A1},$$

$$\varphi_{B1}=\varphi_{B3}=\varphi_{A1},\ \tan\frac{\varphi_{B2}}{2}=\sqrt{2}\tan\varphi_{A1},\ \tan\frac{\varphi_{B4}}{2}=\sin\varphi_{A1},$$

$$\varphi_{C1}=\varphi_{C3}=\varphi_{B3},\ \tan\frac{\varphi_{C2}}{2}=\sqrt{2}\tan\varphi_{A1},\ \tan\frac{\varphi_{C4}}{2}=\sin\varphi_{A1},$$

$$\varphi_{D1}=\varphi_{A2},\ \varphi_{D2}=\varphi_{D6}=\varphi_{A4},$$

$$\cos\varphi_{D4}=\sqrt{2}\sin\varphi_{A2}\sin\varphi_{A4}(\cos\varphi_{A4}-1)+\frac{1}{2}(\cos\varphi_{A2}+1)(\cos^2\varphi_{A4}+1)$$
$$-\cos\varphi_{A2}\sin^2\varphi_{A4}-\cos\varphi_{A2}\cos\varphi_{A4}+\cos\varphi_{A4}-1$$

$$\tan\frac{\varphi_{D3}}{2}=\tan\frac{\varphi_{D5}}{2}=\frac{\cos\varphi_{A2}\cos\varphi_{A4}+\sqrt{2}\sin\varphi_{A2}\sin\varphi_{A4}-\cos\varphi_{A2}-\cos\varphi_{A4}+1}{-\sqrt{2}\sin\varphi_{D4}+\sqrt{2\sin^2\varphi_{D4}+(1-\cos\varphi_{D4})^2-(\cos\varphi_{A2}\cos\varphi_{A4}-\cos\varphi_{A2}-\cos\varphi_{A4}+\sqrt{2}\sin\varphi_{A2}\sin\varphi_{A4}+\cos\varphi_{D4})^2}}.$$

(22)

The input-output curves of dihedral angles and motion sequences (Fig. 11) demonstrate that the origami cube is rigid and flat foldable with one DOF. The folding process of the physical model is shown in Fig. 12, which presents the plane-symmetric folding performance of the origami cube.

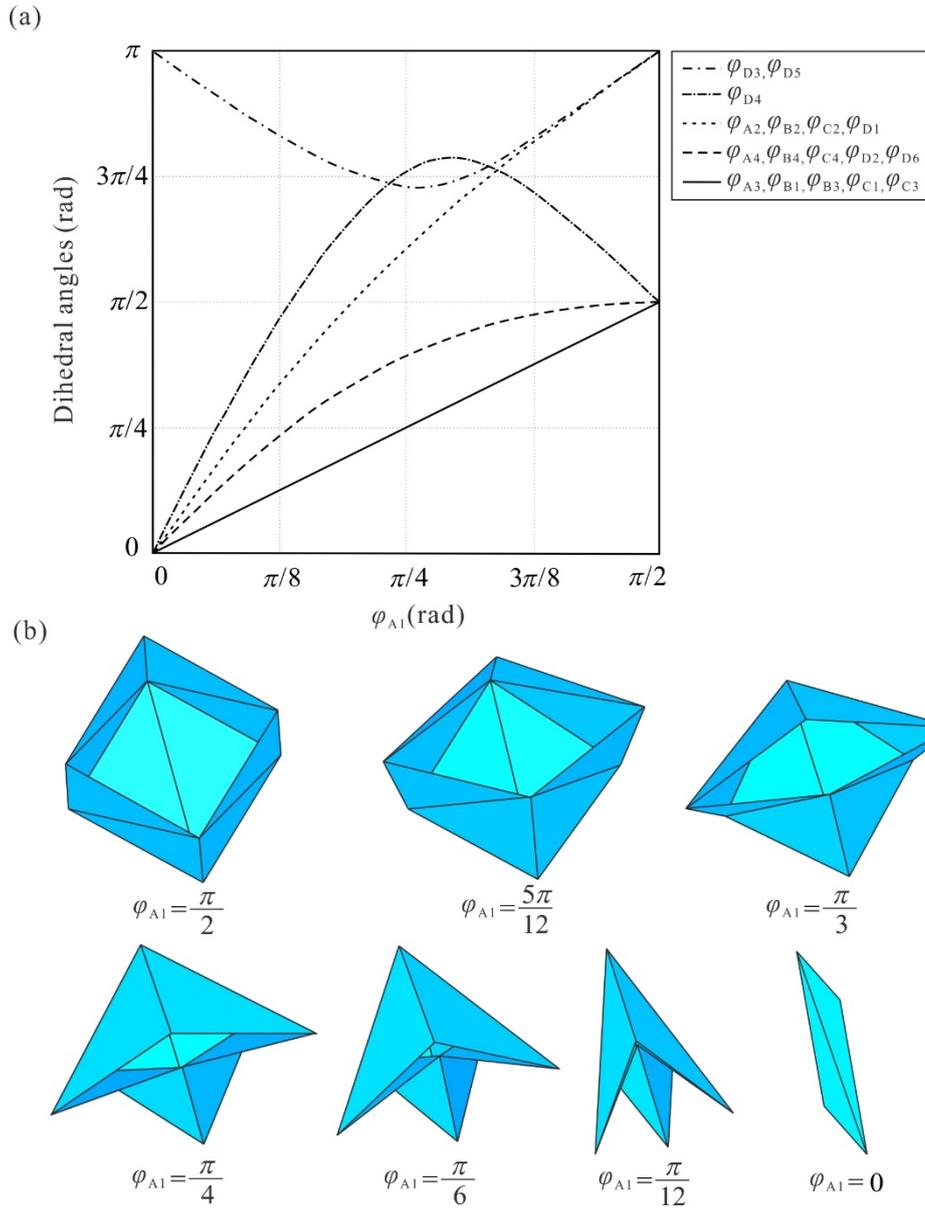

Fig.11. (a) The input-output curves of dihedral angles and (b) the motion sequences of the origami cube in case 3.

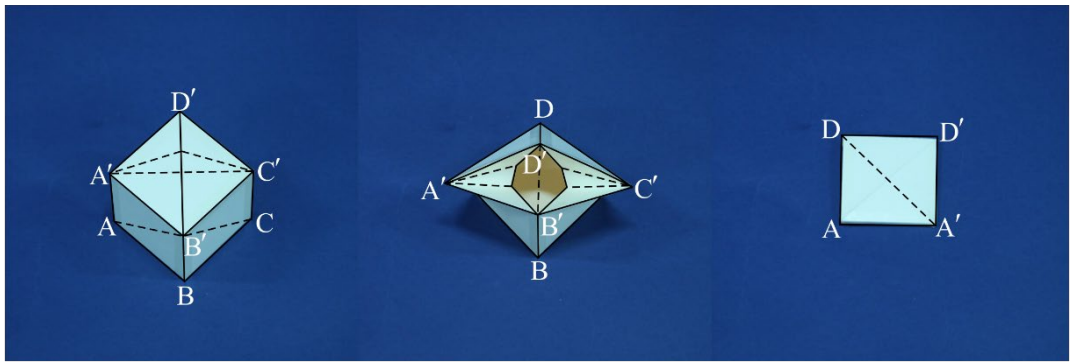

Fig. 12. The folding process of the 4R-4R-4R-6R origami cube.

The pattern ⑨ in case 4 forms a 5R-4R-5R-4R (from vertex A to vertex D) spherical linkage loop (Fig.13) with

$$\alpha_{A12}=\frac{\pi}{2}, \quad \alpha_{A23}=\alpha_{A34}=\alpha_{A45}=\alpha_{A51}=\frac{\pi}{4}, \quad (23a)$$

$$\alpha_{B12}=\alpha_{B23}=\frac{\pi}{4}, \quad \alpha_{B34}=\alpha_{B41}=\frac{\pi}{2}, \quad (23b)$$

$$\alpha_{C12}=\frac{\pi}{2}, \quad \alpha_{C23}=\alpha_{C34}=\alpha_{C45}=\alpha_{C51}=\frac{\pi}{4}, \quad (23c)$$

$$\alpha_{D12}=\alpha_{D23}=\frac{\pi}{4}, \quad \alpha_{D34}=\alpha_{D41}=\frac{\pi}{2}. \quad (23d)$$

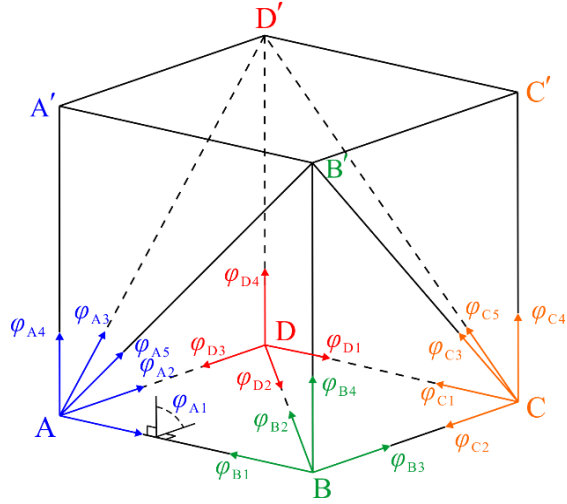

Fig. 13. The crease pattern of case 4 and the 5R-4R-5R-4R spherical linkage loop.

At the bottom of the cube,

$$\varphi_{A1}=\varphi_{B1}, \quad \varphi_{B3}=\varphi_{C2}, \quad \varphi_{C1}=\varphi_{D1}, \quad \varphi_{D3}=\varphi_{A2}, \quad \varphi_{B2}=\varphi_{D2}. \quad (24)$$

Similarly, taking $\varphi_{A1}$ as the input of the linkage loop, $\varphi_{B1}=\varphi_{A1}$ determines the motion of S4R linkage B, while $\varphi_{B2}=\varphi_{D2}$ determines the motion of S4R linkage D. For 2-DOF S5R linkage A, $\varphi_{A1}$ and $\varphi_{A2}=\varphi_{D3}$ are the two inputs. And $\varphi_{C2}=\varphi_{B3}$, $\varphi_{C1}=\varphi_{D1}$ are the two inputs for linkage C. Hence, the kinematic equations of the 5R-4R-5R-4R spherical linkage loop with one DOF are

$$\varphi_{B1} = \varphi_{B3} = \varphi_{A1}, \quad \tan\frac{\varphi_{B2}}{2} = \sqrt{2}\tan\varphi_{A1}, \quad \tan\frac{\varphi_{B4}}{2} = \sin\varphi_{A1},$$

$$\varphi_{D1} = \varphi_{D3} = \varphi_{A1}, \quad \tan\frac{\varphi_{D2}}{2} = \sqrt{2}\tan\varphi_{A1}, \quad \tan\frac{\varphi_{D4}}{2} = \sin\varphi_{A1},$$

$$\varphi_{A2} = \varphi_{A1},$$

$$\tan\frac{\varphi_{A3}}{2} = -\sqrt{2}\left(\sin\varphi_{A1} - \sqrt{\frac{3\sin^2\varphi_{A1} - \cos\varphi_{A1}\sin^2\varphi_{A1} + 2\cos\varphi_{A1} - 2}{1 - \cos\varphi_{A1}}}\right)^{-1},$$

$$\cos\varphi_{A5} = -\sqrt{2}\sin\varphi_{A1}\sin\varphi_{A3} - \cos\varphi_{A1}\cos\varphi_{A3} + \cos\varphi_{A1} - 1,$$

$$\sin\theta_{A4} = \sin\theta_{A1}\cos\theta_{A1}\cos\theta_{A3} + \frac{\sqrt{2}}{2}\sin^2\theta_{A1}\sin\theta_{A3} - \sin\theta_{A1}\cos\theta_{A3},$$

$$\varphi_{Ci} = \varphi_{Ai}, \ (i=1, 2, 3, 4, 5). \tag{25}$$

The 5R-4R-5R-4R origami cube has a twofold plane-symmetric folding performance, the input-output curves of dihedral angles and the motion sequences are shown in Fig. 14. Noted that bifurcation occurs when the bottom square facet of the cube is completely folded, but by subsequent folding, this origami cube can also be folded into a planar square. The folding process of the physical model, in Fig. 15, demonstrates that the 5R-4R-5R-4R origami cube is one-DOF before bifurcation occurs.

## 4. Conclusions and discussion

In this paper, a total of four cases of crease patterns that enable origami cubes rigid and flat foldability with a single DOF has been proposed. The kinematics of four corresponding spherical linkage loops has been studied based on the kinematic equivalence between the rigid origami and the spherical linkages. Due to the different assemblies of spherical linkages, each cube has its own folding performance and symmetric properties. Card models have been fabricated to demonstrate the folding process and validate the study.

For each origami cube in this paper, its foldability derives from the motion of the spherical linkage loop at the bottom of the cube, so the height of origami structure in each case can be optional without considering the top surface. In addition to folding cubes, the new method can be readily extended to the prism structures with diamond bottoms. Furthermore, such kinematic strategies in the rigid origami design can be used to explore the foldability of one-DOF developable polygonal prism structures, which will enhance the engineering applications of origami structures such as foldable cartons, containers, etc.

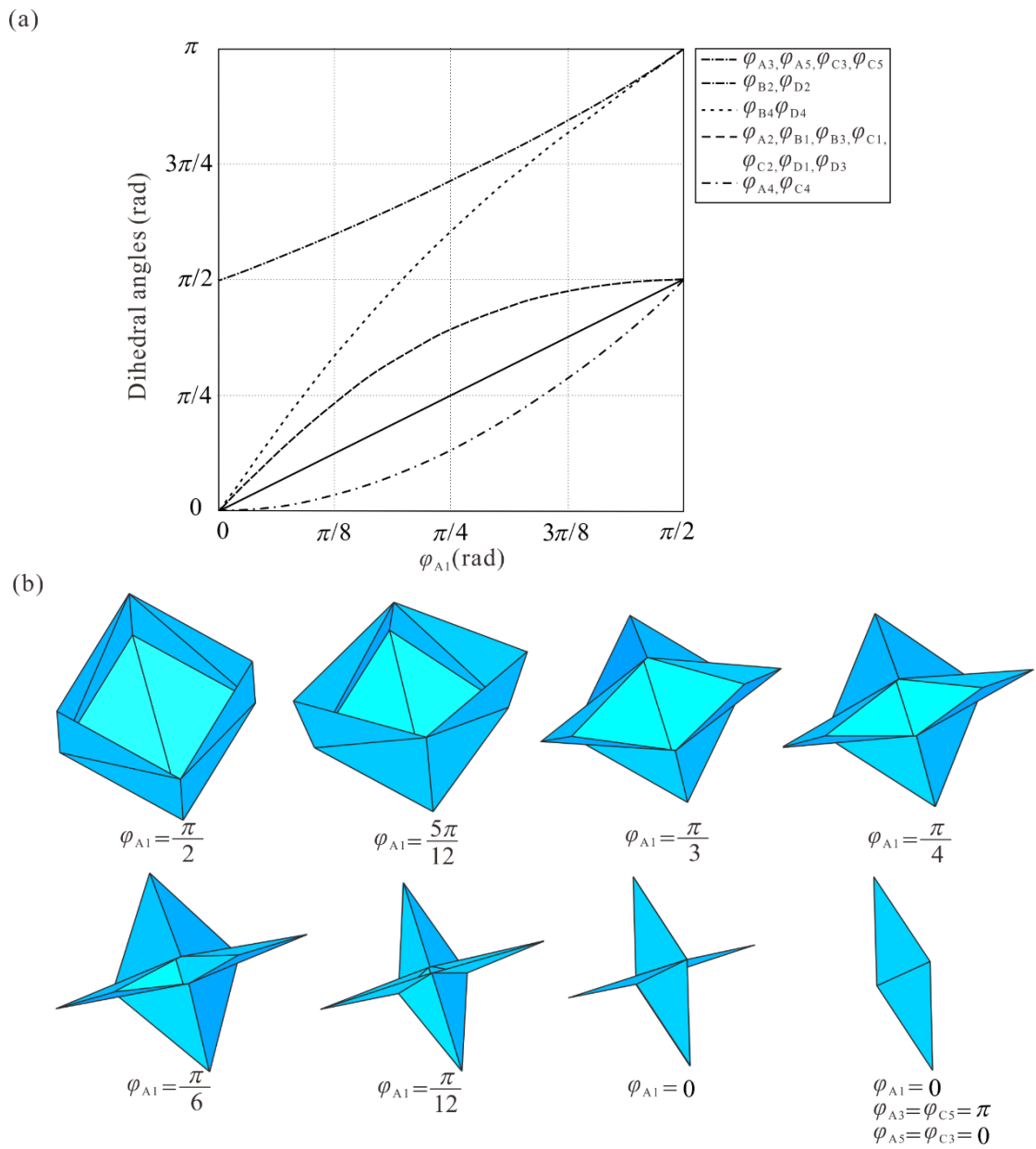

Fig.14. (a) The input-output curves of dihedral angles and (b) the motion sequences of the origami cube in case 4.

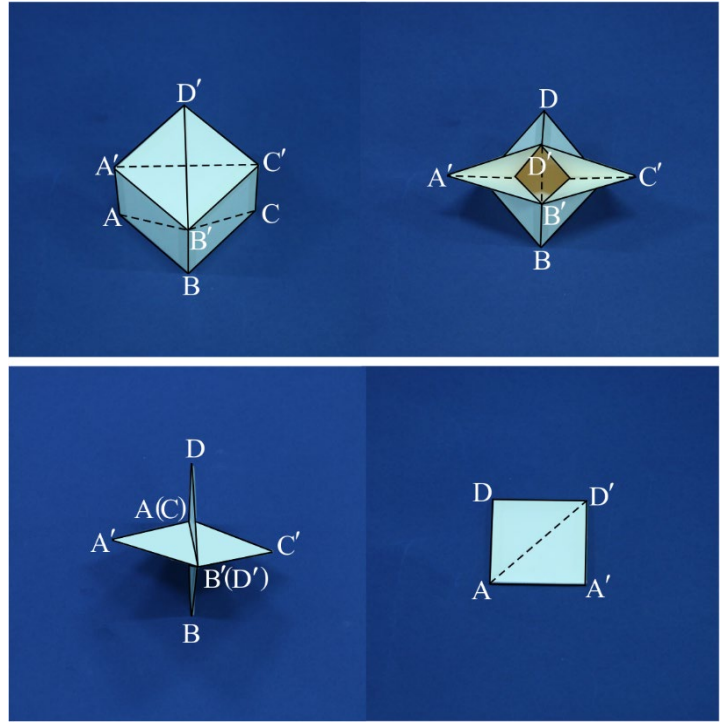

Fig. 15. The folding process of the 5*R*-4*R*-5*R*-4*R* origami cube.

**Acknowledgements**

The authors would like to thank the financial support from the Natural Science Foundation of China (Projects No. 51825503 and No. 51721003).

**References**

Balkcom, D.J., Demaine, E.D., Demaine, M.L., Ochsendorf, J.A., You, Z., 2009. Folding paper shopping bags. In Origami4: Proceedings of the 4th International Meeting of Origami Science, Math, and Education (OSME 2006), 315-334.

Connelly, R., Sabitov, I., Walz, A., 1997. The bellows conjecture. Beitr. Algebra Geom. 38, 1-10.

Dai, J.S., Jones, J. R., 1999. Mobility in metamorphic mechanisms of foldable/erectable kinds. J. Mech. Design 121, 375-382.

Dai, J.S., Cannella, F., 2008. Stiffness characteristics of carton folds for packaging. J. Mech. Design 130, 022305.

Demaine, E.D., O'Rourke, J., 2007. Geometric folding algorithms: linkages, origami, polyhedra. Cambridge university press.

Denavit, J., 1955. A kinematic notation for low pair mechanisms based on matrices. ASME J. Appl. Mech. 22, 215-221.

Guest, S.D., Pellegrino, S., 1994. The folding of triangulated cylinders, Part I: Geometric considerations. J. Appl. Mech. 61, 773-777.

Hull, T. C., 2013. The combinatorics of flat folds: a survey. Proceedings of the Third International Meeting of Origami Science Mathematics & Education, 29-38.

Liu, S., Lv, W., Chen, Y., Lu, G., 2016. Deployable prismatic structures with rigid origami patterns. J. Mech. Robotics, 8, 031002.

Lv, C., Krishnaraju, D., Konjevod, G., Yu, H., Jiang, H., 2014. Origami based mechanical metamaterials. Sci. Rep. 4, 5979.

Miura, K., 1994. Map fold a la Miura style, its physical characteristics and application to the space science. Research of Pattern Formation, 77-90.

Miura, K., 2009. Triangles and quadrangles in space, Symposium of the International Association for Shell and Spatial Structures (50th. 2009. Valencia). Evolution and Trends in Design, Analysis and Construction of Shell and Spatial Structures: Proceedings. Editorial Universitat Politècnica de València.

Nojima, T., 2002. Modelling of folding patterns in flat membranes and cylinders by origami. Jsme International Journal 45, 364-370.

Peraza-Hernandez, E.A., Hartl, D.J., Malak Jr, R.J., Lagoudas, D.C., 2014. Origami-inspired active structures: a synthesis and review. Smart Mater. Struct. 23, 094001.

Schneider, J., 2004. Flat-foldability of origami crease patterns. Swathmore College, December, 10.

Tachi, T., 2009. Generalization of rigid-foldable quadrilateral-mesh origami. Journal of the International Association for Shell & Spatial Structures 50, 173-179.

Tachi, T., 2010. Freeform Rigid-Foldable Structure using Bidirectionally Flat-Foldable Planar Quadrilateral Mesh. Advances in Architectural Geometry 2010, 87-102.

Tachi, T., 2010. Freeform variations of origami. J. Geom. Graph. 14, 203-215.

Wang, K., Chen, Y., 2011. Folding a patterned cylinder by rigid origami. Origami, 5, 265-276.

Wu, W., You, Z., 2010. Modelling rigid origami with quaternions and dual quaternions. Proc. R. Soc. A. 466, 2155-2174.

Wu, W., You, Z., 2011. A solution for folding rigid tall shopping bags. Proc. R. Soc. A. 467, 2561-2574.

Zirbel, S.A., Lang, R.J., Thomson, M.W., Sigel, D.A., Walkemeyer, P.E., Trease, B.P., Magleby, S.P., Howell, L.L., 2013. Accommodating thickness in origami-based deployable arrays. J. Mech. Design 135, 111005.